\documentclass[11pt]{article}
\usepackage{moriond,epsfig}
\usepackage{amsmath,amssymb}
\usepackage{bm}
\usepackage{axodraw}
\usepackage{color}
\usepackage{graphics,graphicx}

\def\be{\begin{equation}}
\def\ee{\end{equation}}
\def\bea{\begin{eqnarray}}
\def\eea{\end{eqnarray}}

\begin{document}
\hspace*{13cm}\vspace{-1cm}DESY 06-068
\vspace*{4cm}
\author{Joerg Jaeckel$^{1}$, Eduard Masso$^{2}$, Javier Redondo$^{2}$, Andreas Ringwald $^{1}$, Fuminobu Takahashi $^{1}$\\[0.3cm]}
\address{$^{1}$ Deutsches Elektronen Synchrotron, Notkestrasse 85,
22607 Hamburg, Germany}
\address{$^{2}$ Grup de F{\'\i}sica Te{\`o}rica and Institut de
F{\'\i}sica d'Altes Energies Universitat Aut{\`o}noma de Barcelona
08193 Bellaterra, Barcelona, Spain}
\title{We need lab experiments to look for axion-like particles}

\maketitle\abstracts{
The PVLAS signal has renewed the interest in light bosons coupled to the electromagnetic field.
However, astrophysical bounds
coming from the lifetime of the sun and the CAST experiment are seemingly in conflict with this result. We discuss effective models that allow
to suppress production of axion-like particles in the sun
and thereby relax the bounds by some orders of magnitude. This stresses the importance of laboratory searches.}

\section{Introduction}
Recently the PVLAS collaboration has reported a rotation of the
polarization plane of an
originally linearly polarized laser beam propagating through a
magnetic field \cite{Zavattini:2005ca}. This signal could be
explained by the existence of a light neutral spin zero boson with a
coupling to two photons, e.g. a pseudoscalar {$\phi$},
\begin{equation}
\label{action1}
{\mathcal{L}}_{\phi}=\frac{1}{2}(\partial_{\mu}\phi)^{2}-\frac{1}{2}m^{2}_{\phi}\phi^{2}
-\frac{g}{4}\phi\tilde{F}^{\mu\nu}F_{\mu\nu},
\end{equation}
with
\begin{equation}
\label{masscoupling} m^{\textrm{PVLAS}}_{\phi}=
(1-1.5)\,\textrm{meV},\quad g^{\textrm{PVLAS}}_{\phi} =(1.7-5)\times 10^{-6}\,\textrm{GeV}^{-1}.
\end{equation}

\begin{figure}
\begin{center}
\scalebox{0.90}[0.90]{
\begin{picture}(190,140)(50,45)
\includegraphics[width=9.5cm]{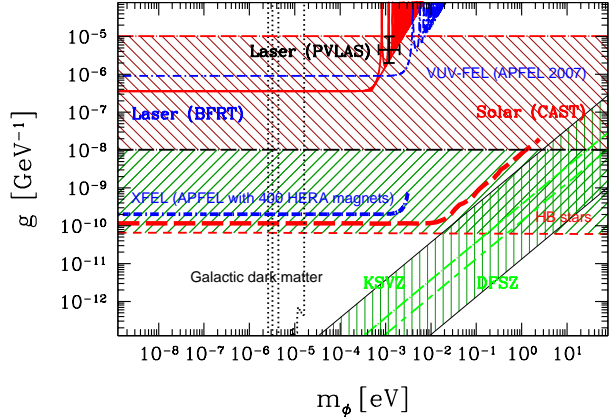}
\end{picture}
}
\end{center}
\caption{Various bounds on the coupling $g$ and mass $m_{\phi}$ of a (light) boson coupled to two photons (areas above single lines are excluded). The green vertically shaded strip gives the range of all reasonable axion models. The two lines within its boundaries give a typical KSVZ and DFSZ model. The green and red diagonally shaded areas give the additional area allowed when we suppress the production of ALP's in the sun (the green smaller one is a little bit more conservative).\hfill \,}
\label{overview}
\end{figure}

The favorite candidate for such a light and neutral particle is
the axion, the pseudo-Goldstone boson of the Peccei-Quinn symmetry
that was proposed to solve the so called strong
CP problem\cite{Peccei:1977hh,Weinberg:1977ma,Wilczek:1977pj}.
However, the PVLAS measurements are not compatible with the
expectations for a standard axion, for which one has a relation that essentially determines
$m_{\phi}g^{-1}_{\phi}$ in terms of QCD quantities.
All natural axion models are
located in the green vertically shaded strip in Fig. \ref{overview}. As we can see from
the same figure, the PVLAS result is far outside this region. Hence, it is probably
not an axion. We will call it an axion like particle (ALP) due to its similar properties.

The troubling point of the particle interpretation of the PVLAS
data is that the action \eqref{action1} with parameters \eqref{masscoupling} already is in conflict
with observations.
Astrophysical considerations based on the production of $\phi$'s from photons
via the coupling $g\phi F\tilde{F}$ in Eq. \eqref{action1}
actually give the strong bounds depicted in Fig. \ref{overview}.
We will
briefly review these bounds in \mbox{Sect. \ref{astrophysics}}.

The motivation for our work\footnote{This note is based on a talk given by J. Jaeckel at the
''Rencontres des Moriond: Contents and structures of the universe'' in La Thuile, Italy in March 2006. For more details see \cite{ourprep}.} is the question: Can we
resolve the conflict between the astrophysical bounds and the particle interpretation of
PVLAS\footnote{For other attempts in this direction see \cite{Masso:2005ym,Jain:2005nh}.} and
how can this be tested?
We attack the first part of this question in Sect. \ref{suppression}. In our final Sect. \ref{conclusions}
we argue that laboratory experiment are the prime tool to give a conclusive answer if the particle
interpretation of PVLAS is invalid or if astrophysical bounds are evaded and PVLAS has detected signals of new
exciting physics.

\enlargethispage{1cm}
\section{Astrophysical bounds revisited}\label{astrophysics}
The basic problem of the particle interpretation of the PVLAS data is that it is in conflict with
astrophysical bounds. For a better understanding of the problem let us briefly review these bounds.
\subsection{Energy loss of stars}

The simplest bound comes from the energy loss argument. If any
weakly interacting particle is produced in a star and
escapes, it takes a certain amount of energy with it, thus
contributing to the stellar luminosity. The amount of energy in
exotic particles can contribute to shorten the duration of the
different phases of stellar evolution, which can be observed (for a review \mbox{see
\cite{Raffelt:1996wa}}).

Here, we focus on the sun, for which we
have a solid standard solar model from which we can accurately calculate
emission of ALPs.

The lifetime of the sun is known to be around $10$ billion years
from radiological studies of radioactive isotopes in the solar
system (see \cite{Bahcall:1995bt}). Solar models
reproduce this quantity (among others). From this one concludes that the exotic contribution
to the luminosity cannot exceed the standard solar luminosity in photons. For our
purposes this means
\begin{equation}
\label{max}
L_{\textrm{ALP}}<L_\odot=3.8\times10^{26}\,\textrm{W} \approx 1.6\ 10^{30}\, \textrm{eV}^2.
\end{equation}
We compute the ALP emission in the standard solar model BP2000 \cite{Bahcall:2000nu} using
Eq. \eqref{action1} and no further assumptions. We obtain a
slightly bigger value than that of \cite{vanBibber:1988ge}, where the calculation was done using
an older solar model.
\begin{equation}
\label{flux}
L_{\textrm{ALP}}= 0.063\, g^2_{10}\ L_\odot,
\end{equation}
where $g_{10}=g\,10^{10}\, \textrm{GeV}$.
Together with Eq. \eqref{max} this gives a bound on the coupling. 
A somewhat strengthened bound\cite{Raffelt:1996wa} including data
from so called horizontal brach (HB) stars is shown in \mbox{Fig. \ref{overview}}.

Another bound comes from the CAST experiment\cite{Beltran:2005ch}. The CAST experiment tries to detect
the axion flux \eqref{flux} by reconverting the axions in the strong
magnetic field generated by an LHC test magnet.
The rate of photons in the detector is
\begin{equation}
\textrm{rate}\sim g^{2} L_{\textrm{ALP}}\sim g^4.
\label{CAST}
\end{equation}
So far no significant photon flux has been measured. The resulting bound is depicted in Fig. \ref{overview}.
\enlargethispage{1.2cm}
\section{Evading astrophysical bounds}\label{suppression}
Our strategy to evade the astrophysical bounds is rather
simplistic. In the center of the sun where most ALPs are produced
the environment is different from the environment of the PVLAS
experiment. If the parameters $m_{\phi}$ and $M_{\phi}$ depend on
the environment
\begin{equation}
m_{\phi}\rightarrow m_{\phi}(\textrm{environment}), \quad
M_{\phi}\rightarrow M_{\phi}(\textrm{environment}),
\end{equation}
in a suitable way the production of the $\phi$ particles inside
the sun is strongly suppressed. In particular, we consider
the following:
\begin{enumerate}
\item{The density $\rho$ inside the sun is quite high.}
\item{Inside the sun the temperature $T$ is high.} \label{hight}
\item{The average momentum transfer $\langle q\rangle$ in the
Primakoff processes generating the ALP's is high.}
\end{enumerate}

We will not try to construct micro physical explanations but
rather write down simple effective models and fix their parameters
in order to be consistent with PVLAS as well as the astrophysical bounds.

For simplicity we allow in this note only a variation of the coupling $g$.  Suppression via a high mass term in the sun environment is more difficult since it involves a strong coupling to generate the high mass (for more details see \cite{ourprep}).
In addition, we will restrict
the dependence to a single parameter $\alpha=\rho,\, \textrm{T, etc}.$.
We are mainly interested in giving conservative bounds for $g$ and the
suppression scales involved, so instead of guessing possible
dependencies $g=g(\alpha)$ we make the calculations with
the most optimistic suppression, a step function, i.e. if $\alpha>\alpha_{\textrm{crit}}$, $g=0$, and the generation of ALP's in this region is completely suppressed.

The macroscopic quantities $\rho,\, T$ depend more or less
only on the radius. Therefore, we get the following simple
picture. In the center of the sun (where naively most ALP's would
be produced) the suppression is switched off while in the
remaining shell we have no suppression at all. Using this we only
need to calculate the production in the outer shell and compare it
with the production within the whole volume happening in a
scenario without suppression\footnote{For CAST actually one has to
take into account that the CAST detector measures a number and not
an energy flux and is only sensitive in a certain energy range.
This gives a slightly modified suppression factor $\tilde{S}$ (see
\cite{ourprep}).},
\begin{equation}
S(R_{0})=\textrm{suppression factor}=\frac{\textrm{production}(R>R_{0})}{\textrm{production(full sun)}}.
\end{equation}

We treat the emission of ALP's as a small perturbation of the
solar model and therefore can compute the emission of these particles from the
unperturbed solar model.  We have chosen the BP2000 of Bahcall et al
\cite{Bahcall:2000nu}.
The suppression factor  $S$ for some radii is given in Tab. \ref{table1}.

\begin{table}[t]
\begin{center}
\begin{tabular}{|c|c|c|c|}
\hline
 $R_0/R_\odot$ &  $\rho_0/(\textrm{g}\, \textrm{cm}^{-3})$ & $T_0/\textrm{eV}$
 & $S(R_{0})$
 \\
 \hline
 $0$ & $150$ & $1200$
 & $1$ 
 \\
\hline
$0.79$ & $0.1$ & $120$
& $10^{-4}$ 
\\
\hline
$0.97$ & $0.003$ & $12$ 
& $10^{-20}$
\\
\hline
\end{tabular}
\caption{Several values for suppression factors.
\hfill \,}
\end{center}
\label{table1}
\end{table}

Similar reasoning can be applied to $\langle q\rangle$.
However, writing down a model it is preferable to use directly
the microscopic quantity $q$. In this situation one
has to perform the thermal average over
the scattering processes (see \cite{ourprep}).
For the suppression factors in the range $10^{-4}-10^{-20}$ the
resulting critical $q$ lie in the $\textrm{meV}-\textrm{eV}$ range.

Using Eqs. \eqref{flux} and \eqref{CAST} we find
that a suppression
\begin{equation}
S_{\textrm{loss}}=g^{2}_{\textrm{supp}}/g^{2}_{\textrm{loss}},\quad
S_{\textrm{CAST}}=g^{4}_{\textrm{supp}}/g^{4}_{\textrm{CAST}},
\end{equation}
is needed to achieve a less restrictive bound $g_{\textrm{supp}}$.

The necessary critical values for the temperature, density and momentum transfer for certain suppression factors can be read off from Tab. \ref{table1}. It is rather obvious that none of these values is very extreme. However, one has to compare to the values in PVLAS. These are even smaller,
\begin{equation}
\rho_{\textrm{PVLAS}}<2\times 10^{-5} \textrm{g}\,\textrm{cm}^{-3},\quad T<300\,\textrm{K}\approx 0.025\, \textrm{eV},\quad q\sim 10^{-6}\,\textrm{eV}.
\end{equation}
Hence, we have room for some \emph{exotic} possibilities. Even a suppression factor of $10^{-20}$ is marginally possible, allowing for the PVLAS result. This gives the red (large) shaded region in Fig. \ref{overview}. For a somewhat more conservative suppression factor we find the green (smaller) shaded region in Fig. \ref{overview}.

\enlargethispage{1cm}
\section{Conclusions: We need lab experiments!}\label{conclusions}
Naively, the particle interpretation of the PVLAS data is in conflict with astrophysical bounds. If we allow
for an interaction between photons and axion like particles (ALP's) that depends on other physical quantities (density, temperature and momentum are candidates), the production of ALP's can be suppressed and the astrophysical bounds can be evaded. However, the typical scales appearing in these models are rather low (typically eV and smaller) and the physics must be exotic in this sense. Nevertheless, one cannot rule out these exotic possibilities from the start and PVLAS is a good motivation to look more closely. Since astrophysical bounds can be evaded, a true test can only come from laboratory experiments where we have control of the environmental parameters. A conclusive answer about the particle interpretation of PVLAS can come in particular from so called light shining through walls experiments, where the
photon not only disappears but is regenerated. It is exciting that experiments of this type, with enough sensitivity to test PVLAS, could be built in the next one or two years. An example of such an experiment is APFEL 
(Axion Production at a Free-Electron Laser) which has been proposed at DESY (see also  Fig. \ref{overview}) and is sensitive enough to test PVLAS\cite{Rabadan:2005dm}.

\vspace*{0.2cm}
\emph{J. Jaeckel would like to thank the organizers for the wonderful conference.}

\end{document}